\documentclass[12pt,preprint]{aastex}
\pdfoutput=1

\usepackage{psfrag}

\begin{document}

\title{Significant enhancement of ${\rm H_2}$ formation in disk galaxies
under strong ram pressure}

\author{Benjamin Henderson \& Kenji Bekki} 
\affil{
ICRAR,
M468,
The University of Western Australia
35 Stirling Highway, Crawley
Western Australia, 6009, Australia
}

\begin{abstract}

We show, for the first time, that ${\rm H_2}$ formation on dust grains can be enhanced in disk galaxies under strong ram-pressure (RP).  We numerically investigate how the time evolution, of H{\sc i} and ${\rm H_2}$ components in disk galaxies orbiting a group/cluster of galaxies, can be influenced by hydrodynamical interaction between the gaseous components of the galaxies and the hot intra-cluster medium (ICM). We find that compression of H{\sc i} caused by RP increases ${\rm H_2}$ formation in disk galaxies, before RP rapidly strips H{\sc i}, cutting off the fuel supply and causing a drop in ${\rm H_2}$ density. We also find that the level of this ${\rm H_2}$ formation enhancement in a disk galaxy under RP depends on the mass of its host cluster dark matter (DM) halo, initial positions and velocities of the disk galaxy, and disk inclination angle with respect to the orbital plane. We demonstrate that dust growth is a key factor in the evolution of the H{\sc i} and ${\rm H_2}$mass in disk galaxies under strong RP. We discuss how the correlation between ${\rm H_2}$ fractions and surface gas densities of disk galaxies evolves with time in the galaxies under RP. We also discuss whether or not galaxy-wide star formation rates (SFRs) in cluster disk galaxies can be enhanced by RP if the SFRs depend on ${\rm H_2}$ densities.
\end{abstract}


\section{Introduction}

It is well known that galaxy environments play key roles in 
morphological transformation of galaxies,
enhancement and quenching of galaxy-wide star formation,
and cold gas evolution of galaxies
(e.g., Dressler 1980; Koopmann \& Kenny 2004;
Boselli \& Gavazzi 2006; Bundy et al. 2010; Villalobos et al. 2014; 
Pasquali 2015; Cortese et al.  2016). 
Dynamical and hydrodynamical interactions influence galaxy evolution
(e.g., Bekki 2009; Bekki \& Couch 2011), mergers and intracluster medium (ICM) affect fundamental galactic parameters with increasing likelihood in higher density regions (e.g., Lambas et al. 2012).
Star forming regions peak in merging systems in small and compact groups, 
increasing SFRs and having important effects on galaxy morphology and evolution. 
Hydrodynamical interactions between ICM and cold interstellar medium
(ISM) of disk galaxies
aid in the quenching of star formation (SF), and hence, reddening of galaxies through `strangulation' (e.g., Peng et al. 2015).

Ram-pressure stripping (RPS) has long been considered to be
one of the most important environmental effects and thus has been
investigated by many observational and theoretical astronomers
(e.g., Gunn \& Gott 1972; Takeda et al. 1984; 
Vollmer et al. 2001;
Kronberger et al. 2008; Book \& Benson 2010; Bekki 2014; McPartland et al. 2016). 
These studies include disk gas stripping (e.g., Roediger \& Hensler 2005), 
halo gas stripping (e.g., Bekki 2009), stripped gas tails (e.g., Irwin et al. 1987; Tonnesen \& Bryan 2012) and 
cold gas compression (e.g., Ebeling et al. 2014).
The compression of cold gas initially induces SF in the galaxy, while the stripping of the disk and halo gas causes an eventual decrease in SF in the galaxy. RP is an essential physical process that controls SF histories and gas evolution in disk galaxies.

These previous studies have not extensively investigated how 
RP of ICM in group and cluster environments
can influence the evolution of H{\sc i} and ${\rm H_2}$ components of 
disk galaxies.
Yagi et al. (2013) showed that SF occurs in the stripped tails from galaxies under RP, stripping of ${\rm H_2}$ from a disk galaxy would 
cause faster SF-quenching in the galaxy as it is the direct fuel in the 
SF process. H{\sc i} stripping only aids in the eventual 
`strangulation' of the galaxy and is a more long 
term effect on the galaxy evolution. 
Compression of the ${\rm H_2}$ left in the disk galaxy induces SF due to an increase in density 
and interactions of ${\rm H_2}$. 
Further work into the RP effect on ${\rm H_2}$ is necessary to 
understand the formation and evolution of the cold gas in disk galaxies
within group and cluster environments.

The purpose of this letter is to show, for the first time, that strong RP can significantly enhance ${\rm H_2}$ formation on dust grains in disk galaxies based on the results of numerical simulations. We also show how this enhancement of ${\rm H_2}$ formation depends on model parameters such as orbits within groups/clusters and disk inclination angles with respect to their orbital planes. Since ${\rm H_2}$ is the key parameter in the formation of giant molecular clouds (GMCs) and thus star formation, we discuss the present new results in the context of galaxy-wide ${\rm H_2}$ formation and star formation within dense galaxy environments.
In the present work, we do not discuss the origin of the
shock-excited ${\rm H_2}$ 
in disk galaxies, which has been recently discovered by Wong et al. (2014),
because the present simulation code does not allow us to investigate
${\rm H_2}$ formation through gaseous shocks.

\section{The model}

To investigate time evolution of H{\sc i} and ${\rm H_2}$ content of disk galaxies under ram pressure of ICM, we use our original chemodynamical simulation code that can be run on GPU machines (Bekki 2013, B13). This new code enables us to investigate time evolution of dust and formation of ${\rm H_2}$ on the surface of dust grains in a fully self-consistent manner, though does not include feedback effects of active galactic nuclei (AGNs). 
The most important difference between the present study and previous ones (e.g., Christensen et al. 2012 ; Hopkins et al. 2014) is the self-consistent time evolution of dust abundances deriving H2 abundances used in the present study, whereas others used constant dust-to-metal ratios (i.e., no evolution of dust abundances).
In the present study, a disk galaxy moves in a cluster with a halo dark matter (DM) mass, $M_{\rm dm}$, and an ICM temperature, $T_{\rm ICM}$, allowing its cold gas to be influenced by RP of the ICM. The RP force is calculated according to the position and velocity of the galaxy with respect to the cluster centre (for details see Bekki 2014, B14).

The disk galaxy is composed of dark matter halo, stellar disk, stellar bulge and gaseous disk, similar to the Milky Way (MW-type). The total masses of dark matter halo, stellar disk, gas disk, and bulge are set to be $10^{12} M_{\odot}$, $6 \times 10^{10} M_{\odot}$, $6 \times 10^{9} M_{\odot}$ and $10^{10} M_{\odot}$, respectively. The mass ratio of dark matter halo to disk is fixed at 16.7. We adopt the density distribution of the NFW halo (Navarro, Frenk \& White 1996) suggested from CDM simulations, with $c$-parameter and virial radius set to 10 and 245 kpc, respectively. The spherical bulge is represented by the Hernquist model with scale length of 0.7 kpc and size of 3.5 kpc. The radial ($R$) and vertical ($Z$) density profiles of the stellar disk are assumed to be proportional to $\exp (-R/R_{0}) $ with scale length $R_{0}=0.2R_{\rm s}$ and to ${\rm sech}^2 (Z/Z_{0})$ with scale length $Z_{0}=0.04R_{\rm s}$, respectively. The gas disk, where $R_{\rm g}=2R_{\rm s}$, has radial and vertical scale lengths of $0.2R_{\rm g}$ and $0.02R_{\rm g}$, respectively. In the present MW-type models the exponential disk has $R_{\rm s}=17.5$ kpc and $R_{\rm g}=35$ kpc and a Toomre's parameter of $Q=1.5$. Star formation, chemical evolution, and dust evolution are included in exactly the same way as in B13. The total number of particle used in the MW-type model is 1033400, with gravitational softening length of 2.1 kpc and 200 pc for the disk dark and baryonic components, respectively.
The maximum hydrogen density in the present simulation is $\sim 8 \times 10^3$ cm$^{-3}$, which is significantly lower than ${\rm H_2}$ density of molecular clouds (MCs) cores ($\sim 10^5$ cm$^{-3}$). This lack of resolution for MCs cores will not be a problem in this study, because we focus exclusively on galaxy-scale ${\rm H_2}$ distribution.

To avoid a large number of gas particles to represent the entire ICM (e.g., Abadi et al. 1999), it is represented by $60^3$ SPH particles in a cube with the size of $R_{\rm ICM}$ and a disk galaxy is placed in the exact centre. The initial velocity of each SPH particle for the ICM is set to be ($V_{\rm r}(t=0)$, 0, 0), where $V_{\rm r}$ is determined by the position of the galaxy with respect to the cluster centre. The orbit direction is chosen as the $x$-axis and disk inclination angles $\theta$ and $\phi$ are defined with respect to the $x$-$y$ plane of the ICM cube. $\theta$ is the angle between the spin axis of the disk and the $z$-axis (i.e., $\theta=0$ means the spin axis parallel to the $z$-axis), whereas $\phi$ is the angle between the $x$-$y$ projected spin axis and the $x$-axis.

The mass of ICM particle is time-dependent and calculated according to the mass density of ICM at the position of the galaxy at each time step whereas $T_{\rm ICM}$ is assumed to be constant. The ICM is assumed to have a uniform distribution within the cube with $R_{\rm ICM}$ set to be $6R_{\rm s}$ and $T_{\rm ICM}$ set to be $5.6 \times 10^6$K and $2.6 \times 10^7$ K for $M_{\rm h}=10^{13} M_{\odot}$ and $M_{\rm h}=10^{14} M_{\odot}$, respectively; $T_{\rm ICM}$ is much higher than the temperature of cold ISM in galaxies. We include periodic boundary conditions (at $R_{\rm ICM}$) for the ICM SPH particles leaving the cube. We mainly described the results of 10 representative models in the present study, because these models show typical behaviours of ${\rm H_2}$ evolution in disks under strong RP and intriguing parameter dependence, see Table 1 for model parameter values.

\section{Results}

Figure 1 shows how RP influences the time evolution of 
H{\sc i} and ${\rm H_2}$ masses and spatial distributions 
in a disk galaxy for five models (M1-M5). 
In the mass distributions of model M1,
it can be seen that gas is first strongly compressed and then stripped efficiently form the galaxy. 
At $T=1.41$ Gyr, just before peri-centre passage,
the cold gas has clumped together with small amounts stripped.
The majority of the stripping can in fact be seen at $T=1.76$ Gyr, 
corresponding to the galaxy leaving the cluster central region. 
This indicates that the gas in the disk galaxy first undergoes 
compression (a shallower decrease in H{\sc i} mass) and is then 
stripped (a steeper decrease in H{\sc i} mass). 
Compression of the H{\sc i} can enhance H{\sc i} density
significantly, particularly in the central region,
and thus drives efficient formation of ${\rm H_2}$
on dust grains in the disk galaxy.
As a results, ${\rm H_2}$ mass can significantly increase
with a peak occurring at the peri-centre passage.
${\rm H_2}$ mass begins to decrease after this 
through stripping of gas and formation of new stars from ${\rm H_2}$. 
The initial decrease of ${\rm H_2}$ mass at $T<0.5$ Gyr is due to
the rapid gas consumption by SF.

As shown in Figure 1,
RP effects on ${\rm H_2}$
differ slightly for the five models (M1-M5) with different disk
inclination angles 
($\theta$ and $\phi$).
Although ${\rm H_2}$ formation enhancement can be seen in all
of these models, the level of this enhancement depends weakly on
$\theta$ and $\phi$.
The more edge-on models (M1, M2 and M5) 
have a higher rate of ${\rm H_2}$ loss, 
through ${\rm H_2}$ stripping and star formation 
(driven by compression of ${\rm H_2}$). 
This indicates ${\rm H_2}$ formation efficiency on dust grains in disks under RP depends on the inclination of the disk to the orbital plane of the galaxy.
The models M1 and M5 have higher gas mass loss and RP effects on 
the disk galaxy than the edge-on model M2, which is an interesting result suggesting there is a unique 
angle where RP has most effect. However, 
this is beyond the scope of this letter and we will discuss the
origin of this further in future papers.

Figure 2 describes how the correlation between gas surface density
($\Sigma_{\rm g}$)
and ${\rm H_2}$ fraction can change with time in the disk galaxy under RP
for the fiducial model M1.
In the early phase of disk galaxy evolution ($T=0.35$ Gyr),
the galaxy shows a weak positive correlation between $\Sigma_{\rm g}$
and $f_{\rm H_2}$ for $\Sigma_{\rm g} \le 10$ $M_{\odot}$ pc$^{-2}$.
There is an increase in ${\rm H_2}$ gas fraction between $T=0.35$ Gyr
and $T=1.41$ Gyr in local gaseous regions,
indicating ${\rm H_2}$ formation is enhanced in these local
regions of the galaxy. 
There is little stripping of gas between
$T=0.35$ and 1.41 Gyr, as overall surface gas density does not change 
so significantly. 
Stripping of gas is evident between $T=1.41$ Gyr and $T=1.76$ Gyr 
by the large drop in the overall surface gas density. 
These results are consistent with and reinforce what is shown in 
Figure 1. The move to low mean surface gas density 
and high ${\rm H_2}$ fraction in individual gaseous regions
of the disk galaxy at $T=2.47$ Gyr
is indicative of H{\sc i} being stripped throughout the galaxy 
and not just the outer disk. The almost horizontal line of average ${\rm H_2}$
fraction indicates that H{\sc i} and ${\rm H_2}$ are 
confined to the same pixels (i.e., the same local regions) 
within the disk galaxy.

Figure 3 demonstrates how different galaxy orbits within
clusters of galaxies and different cluster masses ($M_{\rm dm}$) influence
time evolution of H{\sc i} and ${\rm H_2}$ mass in disk galaxies.
The peak of ${\rm H_2}$ mass still corresponds to the peri-centre
passages of galaxies in these models (M6-M9). 
Model M6 has a shallower orbit than M1-M5, due to a 
higher initial galactic velocity, causing less RP to
influence the disk galaxy and less gas to be stripped.
${\rm H_2}$ mass peaks in this model significantly
shallower than models with smaller peri-centre distances (M1-M5).
Similarly to models M1-M5 in Figure 1,
the most prominent stripping occurs as the galaxy leaves the cluster central region. 
Model M7, with lower $M_{\rm dm}$ (group-like halo) and an 
effective tighter orbit of the galaxy,
produces more rapid stripping of H{\sc i} (steep slope in H{\sc i} mass) 
resulting in less ${\rm H_2}$ formation (low ${\rm H_2}$ mass peak) 
than other models (M1-M5). This is likely due to 
the disk galaxy spending more time in the most central region of the 
group where ICM is higher density. 
Contrast to this, model M8 has a wide orbit 
never passing the most central region of the cluster. 
There is little RPS of H{\sc i} and no peak in ${\rm H_2}$ mass, 
as H{\sc i} is not being compressed and not driving formation of ${\rm H_2}$. 
Model M9, with a very close orbit (due to an initial position closer to the cluster centre) and a lower $M_{\rm dm}$ (group-like halo), 
has two peri-centre passes during $\sim$2.8 Gyr. H{\sc i} is rapidly stripped due to the galaxy being in the cluster central region and ${\rm H_2}$ is formed and stripped during the first peri-centre passage. 
During the second passage little increase in ${\rm H_2}$ mass is seen,
as there is insufficient H{\sc i} to compress and form copious
amounts of ${\rm H_2}$.

Figure 4 shows how dust growth can influence 
time evolution of ${\rm H_2}$ and SF histories in disk galaxies under RP.
${\rm H_2}$ mass in the model M10 without dust physics
is considerably lower with only a slight peak at the peri-centre pass. This indicates that 
little ${\rm H_2}$ formation is occurring in the disk galaxy for this model
(M10). Owing to consumption of gas and dust by star formation,
${\rm H_2}$ formation on dust grains cannot continue to be effective
without dust mass
increase through accretion of gas-phase metals onto dust grains (i.e.,
dust growth). It is interesting to note that appreciable enhancement of galaxy-wide star formation 
can be seen in these two models with and without dust physics.
This SF enhancement cannot last long owing to ${\rm H_2}$ consumption
by star formation and gas stripping during and after peri-centre passage.
These results imply that dust physics needs to be included in
modelling of ${\rm H_2}$ evolution of disk galaxies under RP.

\section{Discussion and conclusions}

Formation and evolution of H$_2$ in galaxies is a fundamental 
process in understanding star formation and evolution of 
cold gas in disk galaxies.
We have shown how formation of ${\rm H_2}$ is 
influenced through H{\sc i} compression associated with RP 
of a disk galaxy in the present study.
As the disk galaxy undergoes RP effects
${\rm H_2}$ mass in the galaxy increases, due to enhanced
${\rm H_2}$ formation efficiency --- higher density
of compressed gas
can significantly enhance the conversion from H{\sc i} to ${\rm H_2}$ 
on dust grains.
The present simulations does not have sub-pc resolution so that
the evolution of ${\rm H_2}$ in molecular clouds of
disk galaxies under RPS cannot be investigated.
Using 3D MHD simulations of ${\rm H_2}$ formation within molecular
clouds,
Valdivia et al. (2015) have already investigated (i)
how ${\rm H_2}$ formation in dense molecular clouds
occurs rapidly and (ii) how ${\rm H_2}$ then spreads 
into more defuse regions around the clouds.
It is our future work to investigate how RP can influence
${\rm H_2}$ formation and evolution within individual molecular
clouds in disk galaxy using more sophisticated high-resolution 3D MHD
simulations.

The present study has shown that
there exists a weak positive correlation between
${\rm H_2}$ gas fraction ($f_{\rm H_2}$)
and surface gas density ($\Sigma_{\rm g}$) in disk galaxies under RP
when H{\sc i} is being compressed (yet not stripped).
This result is consistent with observational studies for disk galaxies
in the Virgo cluster (Nakanishi et al. 2006), which have revealed
a similar $f_{\rm H_2}-\Sigma_{\rm g}$ relation
in NGC4254 and NGC4656 (See Figure 3 of Nakanishi et al. 2006).
These two galaxies are further from the cluster centre 
and are concluded by Nakanishi et al. (2006) as to have not yet been stripped of H{\sc i}. 
The other three galaxies in their paper (NGC4402, NGC4569 and NGC4579) are 
stripped of H{\sc i} and the gas fraction and surface density 
are consistent with our findings after strong RPS.
So far only a few observational studies have investigated
a $f_{\rm H_2}-\Sigma_{\rm g}$ relation of
disk galaxies in cluster environments.
Future observations of H{\sc i}
and ${\rm H_2}$ of disk galaxies in clusters will allow us to
discuss the possible influence of RP on ${\rm H_2}$ formation and
evolution in disk galaxies under RP.

The present study has revealed that the evolution
of ${\rm H_2}$ in disk galaxies under RP
depends strongly on their orbits, disk inclination angles, and cluster/group masses
($M_{\rm dm}$). For example, models with a more edge-on disk galaxy orientation show
more efficient ${\rm H_2}$ formation by compression from RP.
Also, in a lower DM mass cluster halo, the disk galaxy is required to 
have a closer orbit to the cluster centre for RP to be effective, 
as more time needs to be spent in higher density ICM region. 
These results imply that SFHs of these galaxies can be
significantly influenced by these parameters, because galaxy-wide SFRs can be 
determined by ${\rm H_2}$ densities. 
McPartland et al. (2016) have recently discussed
how a galaxy falling into a cluster at high velocities
causes an extreme RPS event and gives rise to the `jellyfish' morphology classification (Ebeling et al. 2014) and would possibly have higher
fractions of ${\rm H_2}$ gas.

${\rm H_2}$ formation from gaseous phase H{\sc i} 
is unlikely to occur, as gas particles simply cannot release enough energy when they encounter 
one another (e.g., Gould \& Salpeter 1963; Valdivia et al. 2015). 
The most commonly accepted solution is for ${\rm H_2}$ to form on the 
surface of dust grains in the ISM (e.g., Gould \& Salpeter 1963;
Duley \& Williams 1993; Le Bourlot et al. 2012). 
Here we have shown, for the first time, that dust growth is crucial in
rapid enhancement of ${\rm H_2}$ formation on dust grains
in disk galaxies under RP. 
This implies that large-scale (group/cluster-scale) hydrodynamical effects
can influence very small-scale physics of ${\rm H_2}$ formation on
dust grains.
This also implies that dust evolution can control
the evolution of H{\sc i} and ${\rm H_2}$ content
of disk galaxies in groups/clusters. Almost all previous galaxy-scale and cosmological-scale
simulations of galaxy formation and evolution
in these dense environments did not include dust physics (e.g., Bekki 2013),
and accordingly their predictions on cold gas content would not
be so robust.
It is thus our future study to understand better the important roles of dust in
evolution of cold gas of group/cluster
galaxies using cosmological simulations
with more sophisticated models of dust evolution.

\acknowledgments
We are grateful to the anonymous referee  for constructive and
useful comments.

\begin{deluxetable}{llllllll}
\footnotesize  
\tablecaption{Description of parameter values
for the ten representative models.
\label{tbl-1}}
\tablewidth{-2pt}
\tablehead{
\colhead{  Model ID  \tablenotemark{a} } &
\colhead{  $M_{\rm dm}$  \tablenotemark{b} } &
\colhead{  $X_{\rm g}$  \tablenotemark{c}} &
\colhead{  $Y_{\rm g}$  \tablenotemark{d}} &
\colhead{  $V_{\rm g}$ \tablenotemark{e} } & 
\colhead{  $\theta$  \tablenotemark{f} }  &
\colhead{  $\phi$  \tablenotemark{g} }  &
\colhead{  Dust growth  \tablenotemark{h} }  }
\startdata
M1 & 1.0 & 0.5 & 0.5 & 0.3 & 45 & 30 & Yes  \\
M2 & 1.0 & 0.5 & 0.5 & 0.3 & 0 & 0 & Yes  \\
M3 & 1.0 & 0.5 & 0.5 & 0.3 & 90 & 0 & Yes  \\
M4 & 1.0 & 0.5 & 0.5 & 0.3 & 60 & 30 & Yes  \\
M5 & 1.0 & 0.5 & 0.5 & 0.3 & 30 & 0 & Yes  \\
M6 & 1.0 & 0.5 & 0.5 & 0.5 & 45 & 30 & Yes  \\
M7 & 0.1 & 0.5 & 0.5 & 0.3 & 45 & 30 & Yes  \\
M8 & 1.0 & 1.0 & 1.0 & 0.3 & 45 & 30 & Yes  \\
M9 & 0.1 & 0.3 & 0.3 & 0.3 & 45 & 30 & Yes  \\
M10 & 1.0 & 0.5 & 0.5 & 0.3 & 45 & 30 & No  \\
\enddata
\tablenotetext{a}{
The fiducial model is M1.
}
\tablenotetext{b}{
The initial group/cluster dark matter halo mass in units of $10^{14} M_{\odot}$.
}
\tablenotetext{c}{
The initial $x$-position in units of $r_{\rm vir}$, where $r_{\rm vir}$
is the virial radius of the group/cluster.
}
\tablenotetext{d}{
The initial $y$-position in units of $r_{\rm vir}$, where $r_{\rm vir}$
is the virial radius of the group/cluster.
}
\tablenotetext{e}{
The initial $x$-component of the velocity in units of $v_{\rm c}$, where $v_{\rm c}$ 
is the circular velocity at the position.
}
\tablenotetext{f}{
The initial inclination angle $\theta$ of a disk galaxy.
}
\tablenotetext{g}{
The initial inclination angle $\phi$ of a disk galaxy.
}
\tablenotetext{h}{
`Yes' (`No') means dust growth and destruction
are (not) included in the model.
}
\end{deluxetable}

\begin{figure}
\epsscale{0.98}
\plotone{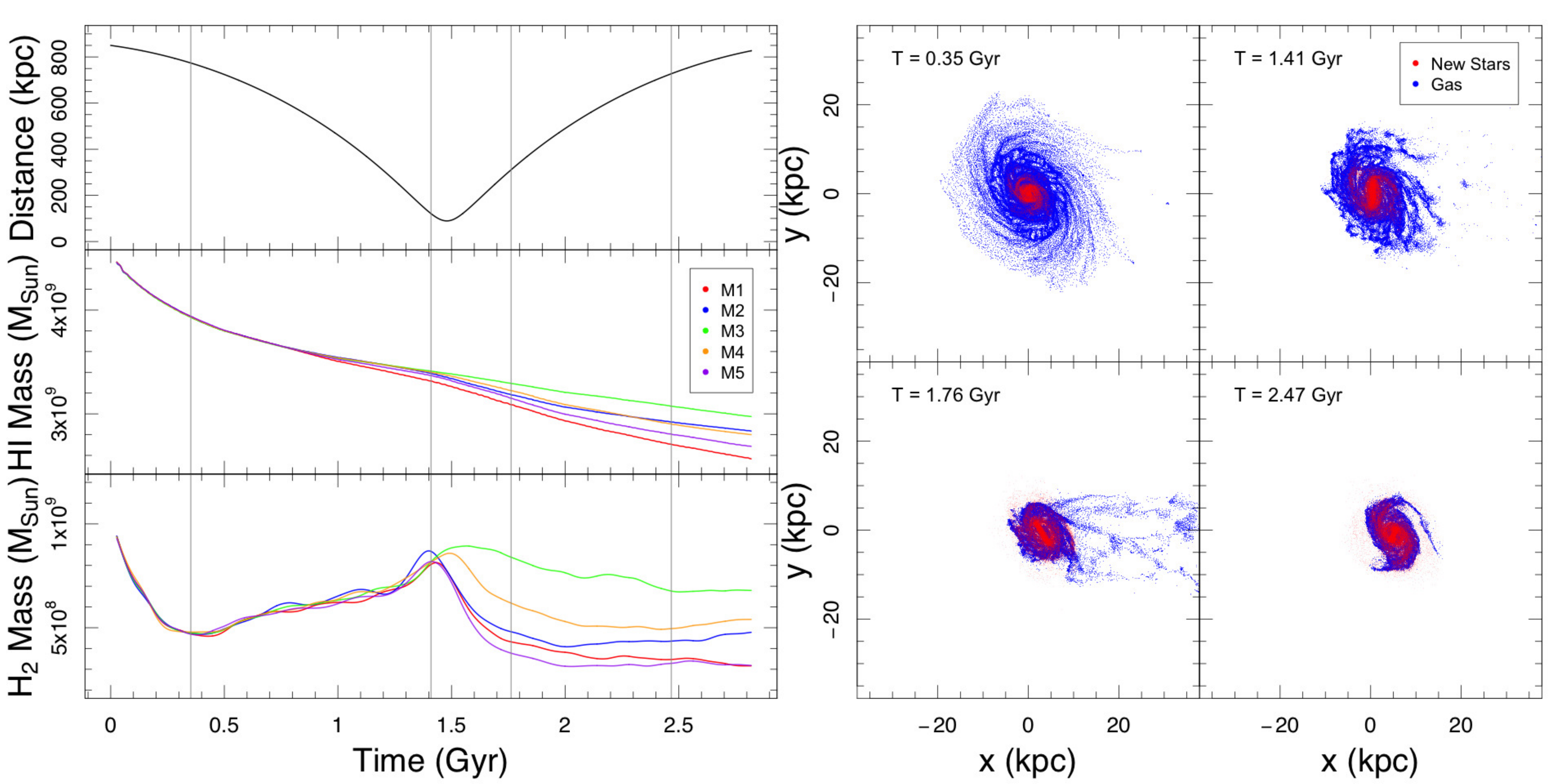}
\figcaption{
Left is the galaxy orbit (top), H{\sc i} mass (middle) and the ${\rm H_2}$ mass (bottom) over $\sim$2.8 Gyr evolution of disks for the models M1, M2, M3, M4 and M5 (red, blue, green, orange and purple respectively). The orbit is represented as the distance of the galaxy to the centre of the cluster in kpc. ${\rm H_2}$ and H{\sc i} mass are measured in solar masses. Right is the positions (in units of kpc) of the new star and gas particles (red and blue respectively), for model M1, projected onto the x-y plane showing four time steps ($T=0.35$ Gyr, $T=1.41$ Gyr, $T=1.76$ Gyr, and $T=2.47$ Gyr), indicated on the left by vertical grey lines.
\label{fig-1}}
\end{figure}

\begin{figure}
\epsscale{0.98}
\plotone{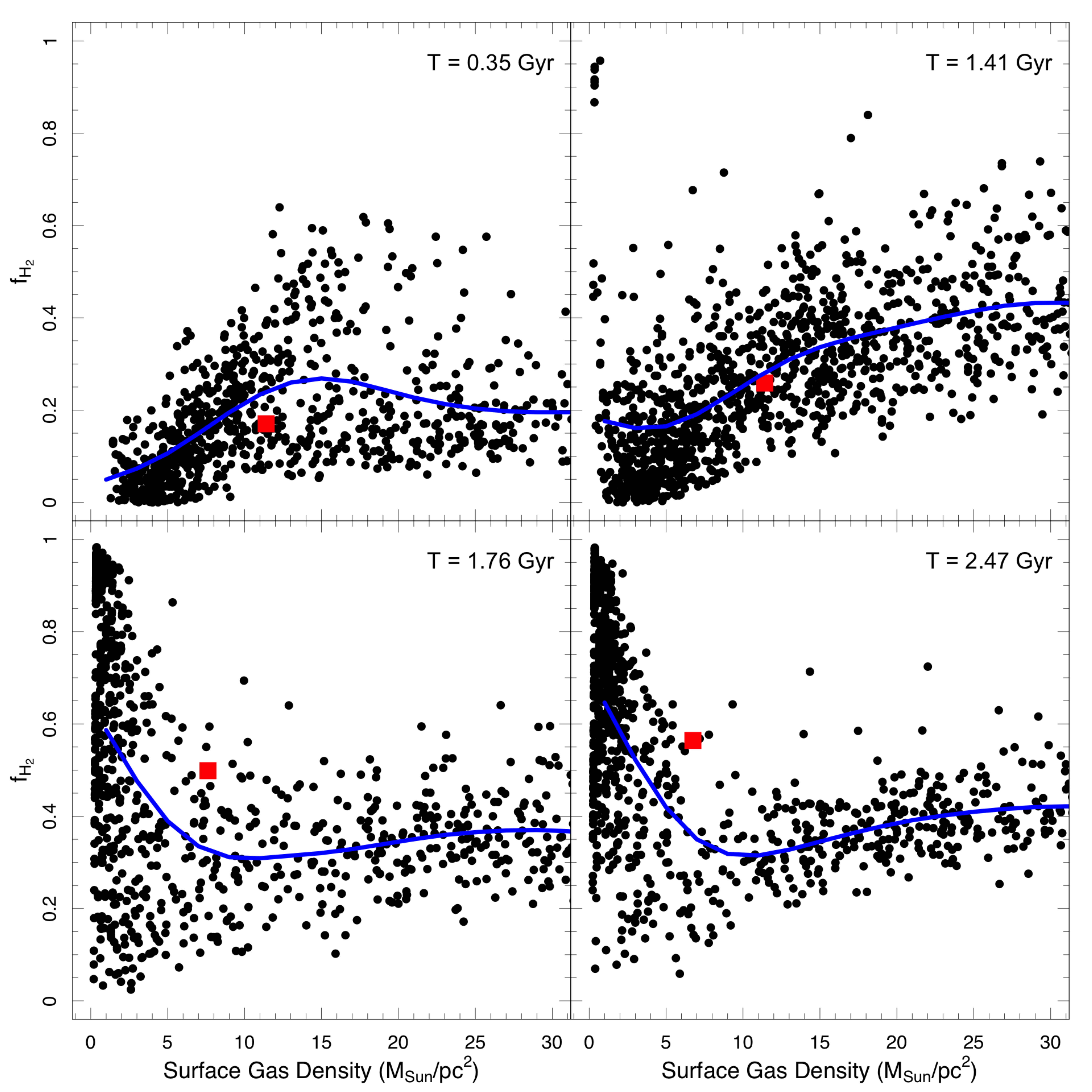}
\figcaption{
${\rm H_2}$ gas fraction and surface gas density are determined for a 50 by 50 grid points in the $x$-$y$ plane for model M1, 25 kpc$^2$ from the centre. Points represent values for one grid point. ${\rm H_2}$ gas fraction is ${\rm H_2}$ mass over H{\sc i} mass and surface density is total gas mass over the area of the pixel in units of $M_{\odot}$/pc$^2$. The red squares are the average gas fraction and the average surface density, and the blue lines are the average gas fraction over the surface density.  Four time steps are shown: $T=0.35$ Gyr (top left), $T=1.41$ Gyr (top right), $T=1.76$ Gyr (bottom left), and $T=2.47$ Gyr (bottom right).
\label{fig-2}}
\end{figure}

\begin{figure}
\epsscale{0.98}
\plotone{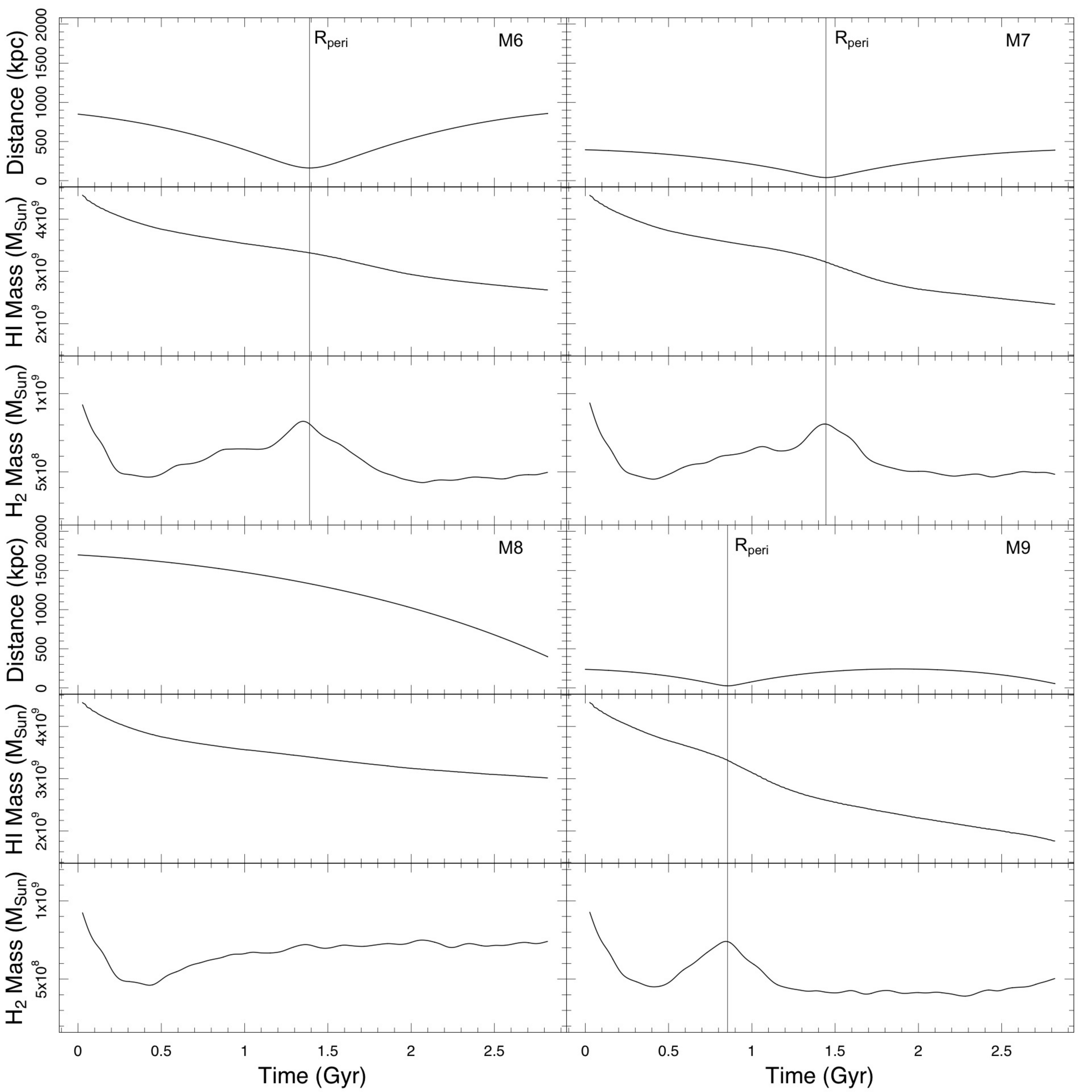}
\figcaption{
Galactic orbit and H{\sc i} and ${\rm H_2}$ mass for the models: M6 (top left), M7 (top right), M8 (bottom left) and M9 (bottom right). The orbit of the galaxy is represented as cluster-centric distance in kpc and ${\rm H_2}$ and H{\sc i} mass are in units of $M_{\odot}$. The vertical line represents the galaxy's peri-centre passage ($\rm R_{peri}$) of the cluster.
\label{fig-3}}
\end{figure}

\begin{figure}
\epsscale{0.98}
\plotone{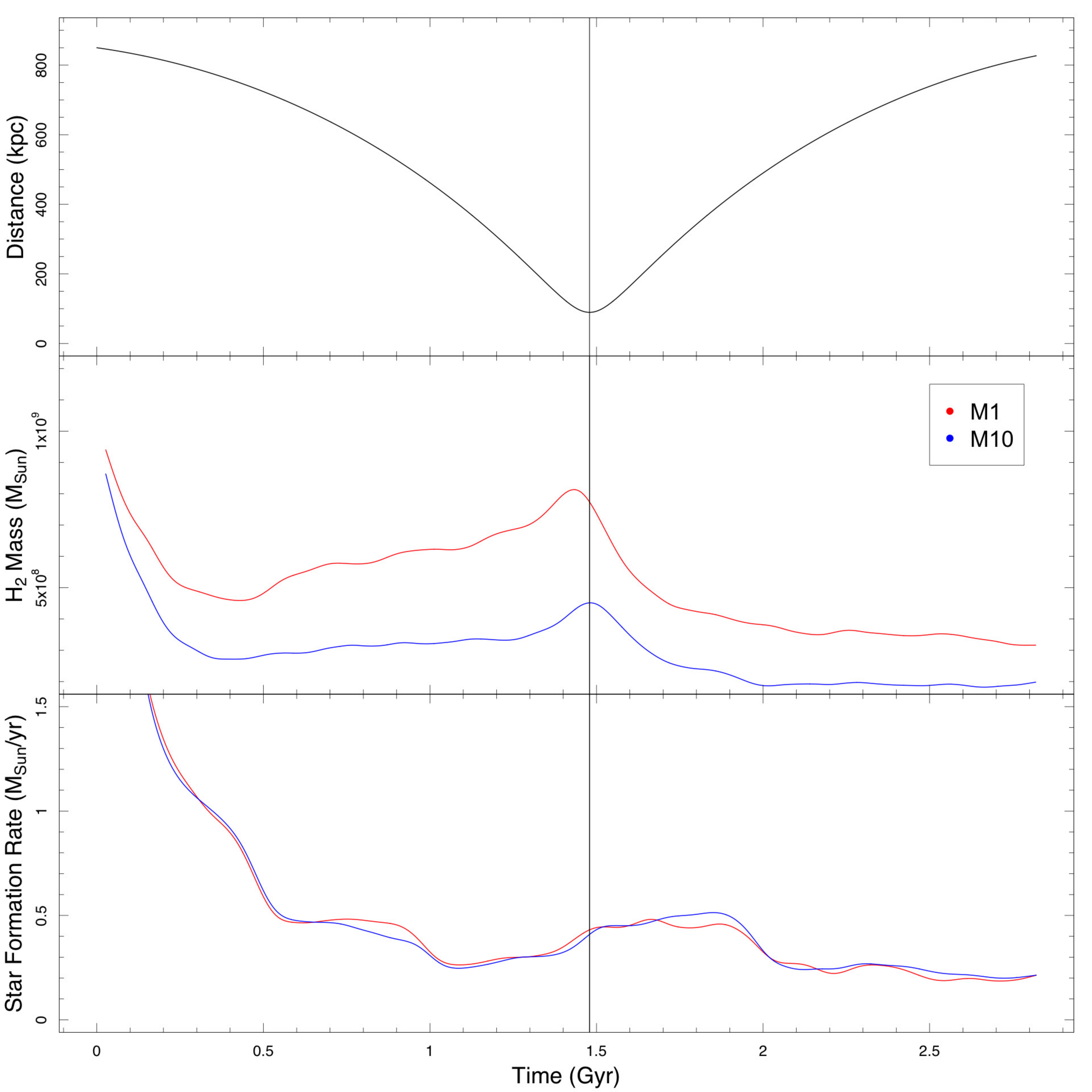}
\figcaption{
The galactic orbit (top), ${\rm H_2}$ mass (middle) and the SFR (bottom) are shown for model M1 (red) and model M10 (blue). As in Figures 1 and 3, orbit is represented as cluster-centric distance in kpc, ${\rm H_2}$ mass is in units of $M_{\odot}$ and SFR is in units of $M_{\odot}$/yr.
\label{fig-4}}
\end{figure}

\end{document}